\UseRawInputEncoding
\documentclass[lettersize,journal]{IEEEtran}
\usepackage{amsmath,amsfonts}
\pdfoutput=1
\usepackage{algorithmic}
\usepackage{array}
\usepackage[caption=false,font=small,labelfont=rm,textfont=rm]{subfig}
\usepackage[font=small,justification=centering,labelfont=rm,textfont=rm]{caption}
\usepackage[linesnumbered,ruled,vlined]{algorithm2e}
\usepackage{textcomp}
\usepackage{stfloats}
\usepackage{url}
\usepackage{multirow}

\usepackage{verbatim}
\usepackage{graphicx}
\def\BibTeX{{\rm B\kern-.05em{\sc i\kern-.025em b}\kern-.08em
		T\kern-.1667em\lower.7ex\hbox{E}\kern-.125emX}}
\usepackage{balance}
\usepackage[utf8]{inputenc}
\usepackage[ruled,vlined]{algorithm2e}
\usepackage{hyperref}
\hypersetup{
	colorlinks = true, 
	linkcolor = blue,
	urlcolor = blue,
	citecolor = blue
}

\begin{document}
	
	\title{Federated Learning Assisted Edge Caching Scheme Based on Lightweight Architecture DDPM}
	
	\author{Xun Li, Qiong Wu,~\IEEEmembership{Senior Member,~IEEE}, Pingyi Fan,~\IEEEmembership{Senior Member,~IEEE}, \\Kezhi Wang,~\IEEEmembership{Senior Member,~IEEE}, Nan Cheng,~\IEEEmembership{Senior Member,~IEEE}, and Khaled B. Letaief,~\IEEEmembership{Fellow,~IEEE}
	\thanks{This work was supported in part by Jiangxi Province Science and Technology Development Programme under Grant 20242BCC32016; in part by the National Natural Science Foundation of China under Grant 61701197; in part by the National Key Research and Development Program of China under Grant 2021YFA1000500(4); in part by the Research Grants Council under the Areas of Excellence Scheme under Grant AoE/E-601/22-R; and in part by the 111 Project under Grant B23008. (Corresponding author: Qiong Wu.)
		
		Xun Li and Qiong Wu are with the School of Internet of Things Engineering, Jiangnan University, Wuxi 214122, China, and also with the School of Information Engineering, Jiangxi Provincial Key Laboratory of Advanced Signal Processing and Intelligent Communications, Nanchang University, Nanchang 330031, China (e-mail: xunli@stu.jiangnan.edu.cn; qiongwu@jiangnan.edu.cn).
		
		Pingyi Fan is with the Department of Electronic Engineering, State Key Laboratory of Space Network and Communications, and the Beijing National Research Center for Information Science and Technology, Tsinghua University, Beijing 100084, China (e-mail: fpy@tsinghua.edu.cn).
		
		Kezhi Wang is with the Department of Computer Science, Brunel University, London, Middlesex UB8 3PH, U.K (e-mail: Kezhi.Wang@brunel.ac.uk).
		
		Nan Cheng is with the State Key Laboratory of ISN and School of Telecommunications Engineering, Xidian University, Xi’an 710071, China (e-mail: dr.nan.cheng@ieee.org).
		
		Khaled B.Letaief is with the Department of Electrical and Computer Engineering, the Hong Kong University of Science and Technology, HongKong (e-mail: eekhaled@ust.hk).}}
	\maketitle

\begin{abstract}
Edge caching is an emerging technology that empowers caching units at edge nodes, allowing users to fetch contents of interest that have been pre-cached at the edge nodes. The key to pre-caching is to maximize the cache hit percentage for cached content without compromising users' privacy. In this letter, we propose a federated learning (FL) assisted edge caching scheme based on lightweight architecture denoising diffusion probabilistic model (LDPM). Our simulation results verify that our proposed scheme achieves a higher cache hit percentage compared to existing FL-based methods and baseline methods.


\end{abstract}

\section{Introduction}
\subsection{Background}
\IEEEPARstart{I}{n} recent years, the surge of smart devices has led to a significant increase in mobile data traffic, putting enormous pressure on wireless networks. As users become increasingly reliant on user devices such as smartphones and computers to access content, ensuring satisfactory service quality has become challenging \cite{2}. To address this challenge, edge caching has become an effective solution. By caching user interested content in advance at wireless network edge nodes such as base stations (BS), user can directly obtain requested content from nearby BS instead of remote cloud server. This method can significantly alleviate network congestion, reduce traffic load, reduce service latency, and improve overall system performance \cite{4}. 

However, BSs have limited cache capacity, making it crucial to enhance the cache hit percentage of cached content \cite{cache_2017}. Denoising diffusion probabilistic models (DPMs) have garnered significant attention for their superior generative capabilities, which generate samples through a step-by-step denoising process \cite{ddpm}. Compared to earlier mainstream generative models (e.g., generative adversarial networks (GANs) and Flow-based models), DPMs exhibit more stable training processes and higher sample fidelity. However, the training of DPMs requires significant computational resources and is not suitable for resource-constrained devices. Recently, several lightweight architecture DPMs (LDPMs) have been proposed to address this challenge. In \cite{snap}, Li \textit{et al.} achieved sub-second text-to-image generation on mobile devices for the first time through the design of an efficient U-Net architecture and improved step distillation techniques. In \cite{lightgrad}, Chen \textit{et al.} propose a lightweight DPM suitable for edge devices through a lightweight U-Net architecture design, which only requires 4 denoising steps to generate high-quality speech.

Additionally, model training requires access to users’ personal data. User personal data often contains a lot of privacy sensitive information, and users are unwilling to directly share their data with others, making it difficult to collect and train directly on user data \cite{wwh}. Fortunately, federated learning (FL) can address this issue by enabling the sharing of local models instead of raw user data \cite{FedAvg}. Therefore, it is necessary to introduce FL into edge caching in order to protect user privacy.
\subsection{Related Work}
Currently, there are many studies adopting FL in edge caching. In \cite{FED}, Wang \textit{et al.} combined FL with deep reinforcement learning to achieve decentralized collaborative caching, solving the communication bottleneck of centralized solutions. In \cite{2021cache}, Yu \textit{et al.} conducted edge caching for vehicle environments, integrating mobility prediction with FL to address the issues of high vehicle dynamics and user privacy. However, the aforementioned papers cannot accurately predict the content that users are interested in. The authors in \cite{wwh, zy, cache2024} adopted a “K-nearest neighbor selection mechanism,” calculating similarity based on the interests between users, and using the interest lists of the K most similar “neighbor” users to the target user as auxiliary predictions of the target user’s interests. Although it can predict the content of user interest more accurately, predicting the content of interest to the target user requires the use of the “neighbor” user’s interest list, which to some extent exposes the privacy of the “neighbor” user. In \cite{WGAN}, Wang \textit{et al.} integrated Wasserstein GAN (WGANs) into FL frameworks, enabling accurate prediction without exposing raw user data. However, the training of GAN models demands substantial computational resources, rendering them impractical for deployment on resource-constrained user edge devices.

To the best of our knowledge, no work has been able to achieve high cache hit percentage without leaking user privacy in the resource-constrained user device environment, which is the reason we are writing this letter.

\subsection{Contributions}
In this letter, we propose a FL assisted edge caching scheme based on LDPM \footnote{The source code has been released at: https://github.com/qiongwu86/Federated-Learning-Assisted-Edge-Caching-Scheme-Based-on-Lightweight-Architecture-DDPM}, which achieves a high cache hit percentage without compromising user privacy. The main contributions of our work are as follows:
\begin{itemize}
\item We are the first to combine LDPM with FL for edge caching. Compared to previous federated schemes, our approach achieves higher cache hit percentage without compromising user privacy and is suitable for edge devices.
	
\item To effectively learn the distribution of high-dimensional sparse user data, we use the pre-trained encoder to map the raw user data into a low-dimensional latent space, allowing LDPM to learn the user data distribution in this low-dimensional space.
	
\item To accurately predict the content of interest to users while protecting their privacy, we first propose a federated-based LDPM training algorithm, and then introduce a content popularity prediction method that generates data samples using the global LDPM at the BS to predict the content of interest to users.

\end{itemize}

\section{System Model}
\subsection{System Scenario}
\begin{figure}
	\center
	\includegraphics[scale=0.43]{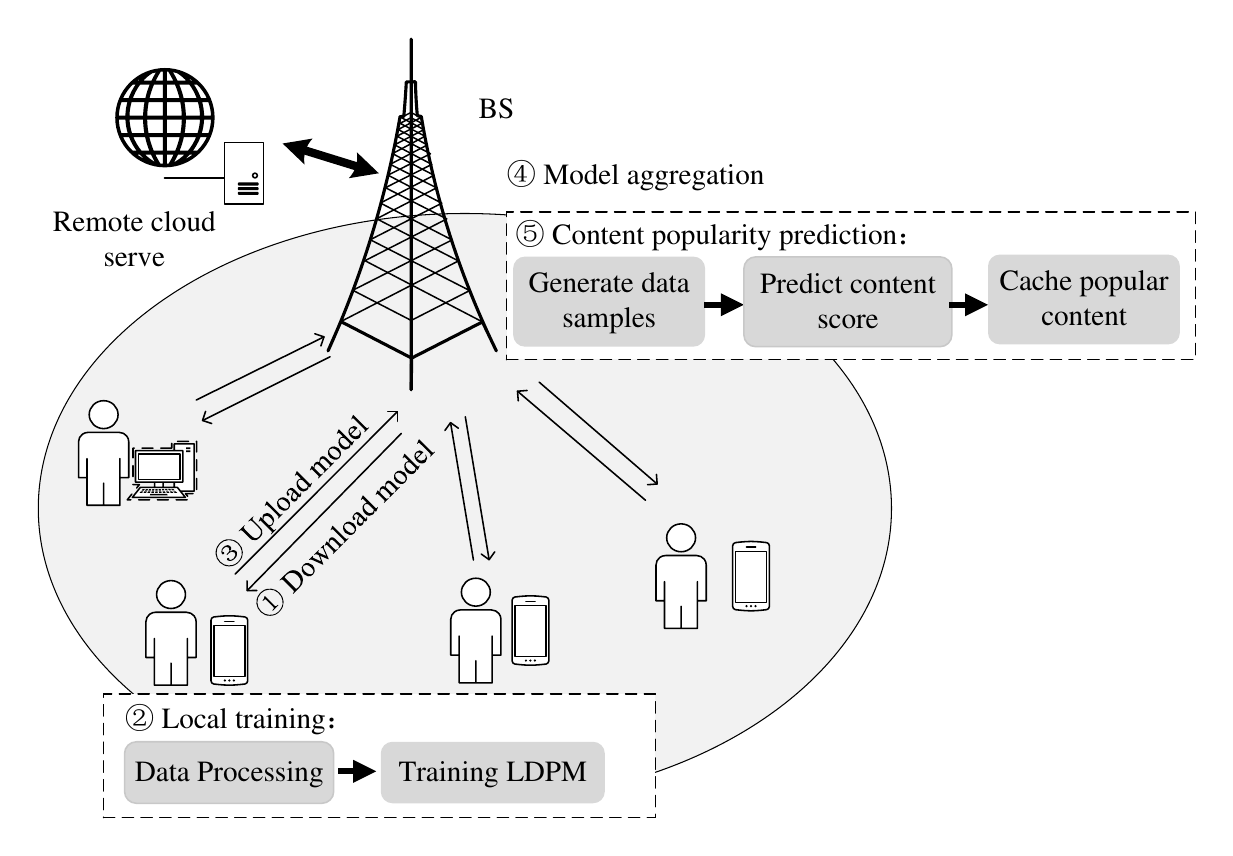}
	\vspace{-0.2cm}
	\caption{System Model.}
	\label{system model}
\end{figure}

The system scenario is shown in Fig. \ref{system model}, where the edge computing network includes a BS, a remote cloud server, and $I$ users. The BS and the remote cloud server are connected via a reliable backhaul link, and users are within the coverage range of the BS. Each user $i = 1, 2, \ldots, I$ has one smart device. The BS is equipped with a caching entity and a decision module. The caching entity has a limited storage capacity and can accommodate up to $N$ contents, while the remote cloud server caches all available content. The decision module acts as a scheduling center for FL and predicts popular contents. When the content requested by a user is cached in the BS, the BS will directly deliver the content to the user. Otherwise, the BS requests the content from the remote cloud server and then delivers it to the user, which results in higher request content delay. Our goal is to maximize the cache hit percentage of user requests by accurately predicting content popularity and proactively caching at the BS.

\subsection{Denoising Diffusion Probabilistic Model (DPM)}
The theoretical foundation of diffusion models stems from the entropy increase-inverse process of non-equilibrium thermodynamic systems. DPM achieves forward diffusion process and reverse diffusion process through parameterized Markov chains.

\subsubsection{Forward Diffusion Process}
By parameterizing a Markov chain with a scheduling strategy $\{\beta_t\}_{t=1}^T$, Gaussian noise is progressively added to the original data, causing the data distribution to gradually perturb towards random noise, where $T$ is the time steps. A single diffusion step can be described as
\begin{equation}
q\left(\mathbf{x}_t|\mathbf{x}_{t-1}\right)=\mathcal{N}\left(\mathbf{x}_t;\sqrt{1-\beta_t}\mathbf{x}_{t-1},\beta_t\mathbf{I}\right).
\end{equation}
In this work, we employ a varying scheduling strategy, where $\{\beta_t\}_{t=1}^T$ increases linearly from $\beta_1=10^{-4}$ to $\beta_T=0.02$. Similar to \cite{chengnan}, we define $\bar{\alpha}_t = \prod_{i=1}^t\alpha_i$ and $\alpha_t = 1-\beta_t$, and can obtain $q\left(\mathbf{x}_t|\mathbf{x}_0\right)$ and $\mathbf{x}_t$ as
\begin{equation}
q\left(\mathbf{x}_t|\mathbf{x}_0\right)=\mathcal{N}\left(\mathbf{x}_t;\sqrt{\overline{\alpha}_t}\mathbf{x}_0,\left(1-\overline{\alpha}_t\right)\mathbf{I}\right),\end{equation}
\begin{equation}
\mathbf{x}_t=\sqrt{\overline{\alpha}_t}\mathbf{x}_0+\sqrt{1-\overline{\alpha}_t}\boldsymbol{\epsilon}_t, \boldsymbol{\epsilon}_t\sim\mathcal{N}\left(0,\mathbf{I}\right).\end{equation}
	
\subsubsection{Reverse Diffusion Process}
By training a neural network $\boldsymbol{\mu}_\theta$ to predict the noise ${\epsilon}_\theta$ at each step, the goal is to recover the original data distribution from the noisy data. The reverse step is parameterized by conditional probability $p_\theta\left(\mathbf{x}_{t-1}|\mathbf{x}_t\right)$, defined as
\begin{equation}
	p_\theta\left(\mathbf{x}_{t-1}|\mathbf{x}_t\right)=\mathcal{N}\Big(\mathbf{x}_{t-1};\boldsymbol{\mu}_\theta\left(\mathbf{x}_t,t\right),\frac{1-\bar{\alpha}_{t-1}}{1-\bar{\alpha}_t}\beta_t\Big),
\end{equation}
\begin{equation}
	\boldsymbol{\mu}_\theta\left(\mathbf{x}_t,t\right)=\frac{1}{\sqrt{\alpha_t}}\Big(\mathbf{x}_t-\frac{\beta_t}{\sqrt{1-\overline{\alpha}_t}}\boldsymbol{\epsilon}_\theta\left(\mathbf{x}_t,t\right)\Big).
\end{equation}

In \cite{ddpm}, Ho \textit{et al.} proposed a simplified objective function for optimization, expressed as
\begin{equation}
\mathcal{L}_{t-1}^{simple}=\mathbb{E}_{t,\mathbf{x}_0,\boldsymbol{\epsilon}\sim\mathcal{N}(0,\mathbf{I})}\Big[\Big\|\boldsymbol{\epsilon}-\boldsymbol{\epsilon}_\theta\Big(\sqrt{\overline{\alpha}_t}\mathbf{x}_0+\sqrt{1-\overline{\alpha}_t}\boldsymbol{\epsilon},t\Big)\Big\|^2\Big].
\label{eq7}
\end{equation}
\vspace{-0.7cm}

\section{Edge Caching Scheme}  
This section introduces the proposed edge caching scheme. We first introduce the federated-based LDPM training algorithm, and then introduce the content popularity prediction algorithm. 

\begin{figure}
	\center
	\includegraphics[scale=0.55]{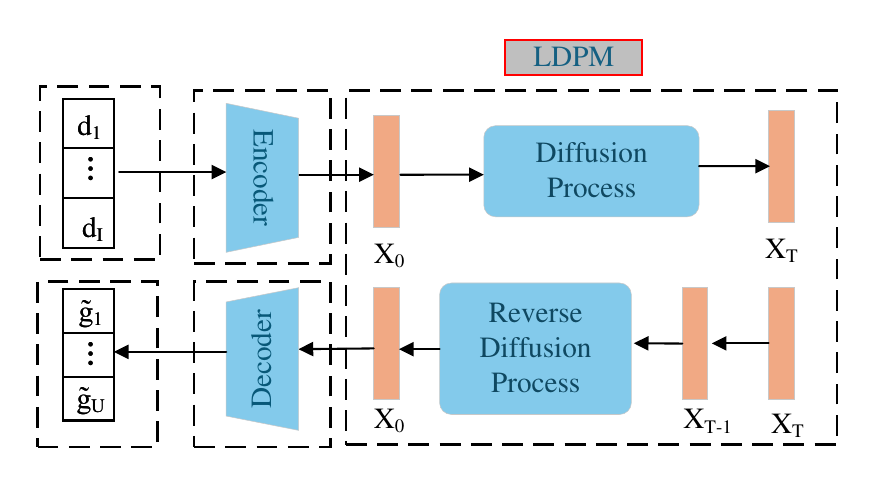}
	\caption{Encoder, Decoder and LDPM framework.}
	\label{system model2}
\end{figure}

Additionally, user data is typically high-dimensional and sparse, causing the Euclidean distances between data points to become uniform, and the noise distribution to become extremely flat \cite{zy}. This makes it difficult for the model to distinguish between signal and noise, and the DPM fails to effectively learn the distribution of user data \cite{laten}. Therefore, we employ pre-trained encoder and decoder to process the data. Before performing local model training, we use the pre-trained encoder to map the raw user data into a low-dimensional latent space, allowing LDPM to learn the user data distribution in this low-dimensional space. Subsequently, when predicting content popularity at the BS, the pre-trained decoder is used to reconstruct the LDPM output data samples back to the original space dimensions. The pre-trained encoder and decoder are trained on the BS using publicly available datasets. Fig. \ref{system model2} illustrates the framework of the encoder, decoder, and LDPM.

\subsection{Federated-Based LDPM Training Algorithm}
During the training process of FL, a total of $R_{\text{max}}$ rounds of training are conducted \cite{FedAvg}. Each round of training $r = 1, 2, \ldots, R_{\text{max}}$ consists of following four steps, corresponding to steps $1-4$ in Fig. \ref{system model}.

\subsubsection{Download Model}
The BS first generates the global LDPM in this step. Let $\omega^r$ represent the global LDPM parameters for the $r$-th round. For the first round of training, the BS initializes global LDPM $\omega^0$. For subsequent rounds, the BS will update the global model at the end of the previous round. Then the BS distributes the global LDPM to users for training.
\subsubsection{Local Training}
The local training process includes data processing and training LDPM. Data processing is mainly used to map the raw user data into a low-dimensional latent space, and then let LDPM learn the distribution of user data in the low-dimensional latent space. 

For iteration \( k \), user $i$ first performs data processing using a pre-trained encoder to map the raw user local data $d_i$ into a low-dimensional latent space \(\hat{d}_i = E(d_{i})\), where \( E(\cdot) \) represents the encoder parameters.

After completing data processing, let LDPM learn the distribution of user local data in the low-dimensional latent space. For iteration \( k \), user \( i \) randomly samples a subset \(\hat{b}_{i,k}^{r}\) from \(\hat{d}_i\). Then, the local loss function for the LDPM can be described as
\begin{equation}
	f(\omega_{i,k}^{r}) = \frac{1}{\left|\hat{b}_{i,k}^r\right|} \sum_{\hat{z} \in \hat{b}_{i,k}^r} L\left(\omega_{i,k}^{r}; \hat{z}\right),
	\label{eq13}
\end{equation}
where $\left|\hat{b}_{i,k}^r\right|$ is the size of the subset \(\hat{b}_{i,k}^{r}\), \( L(\cdot) \) is defined in Equation~\eqref{eq7}, $\hat{z}$ is a data point in $\hat{b}_{i,k}^{r}$, and $\omega_{i,k}^{r}$ refers to the LDPM parameters of user $i$ at the $r$-th round in the $k$-th iteration. Then, the local LDPM is updated as
\begin{equation}
	\omega^{r}_{i,k+1} = \omega_{i,k}^{r} - \eta_d \nabla f(\omega_{i,k}^{r}),
	\label{eq14}
\end{equation}
where \( \eta_d \) is the LDPM learning rate. After completing \( e \) iterations of local LDPM training, the local training process is complete.

\subsubsection{Upload Model}
Each user will upload the locally updated $\omega_i^{r}$ to the BS after completing the local training process.

\subsubsection{Model Aggregation}
After receiving all the local models uploaded by user, BS calculates the weighted sum of models for all users within the coverage area to obtain a new global model,
\begin{equation}
	\omega^{r+1}=\omega^r-\eta \sum_{i=1} \frac{\left|d_{i}\right|}{d} \omega_i^{r},
	\label{eq15}
\end{equation}
where $\left|d_{i}\right|$ is the size of the local data for user $i$, and $d$ is the size of the total data for all users within the BS coverage. So far, the training of federated-based LDPM for the $r$-th round has been finished, and the BS has acquired a new global model $\omega^{r + 1}$. This model will be utilized for the next round of training. Once the number of training rounds reaches $R_{{\text{max}}}$, the entire training process concludes.

\subsection{Content Popularity Prediction}
After completing the training process, the BS uses global LDPM to perform the reverse diffusion process, generating $U$ data samples $g_{u}$, where $U$ is the number of data samples and $u = 1, 2, \ldots, U$. These data samples are fed into the pre-trained decoder to produce reconstructed data samples $\tilde{g}_{u} = D(g_{u})$ in the original data dimensions, where \( D(\cdot) \) represents the dncoder parameters. We use these reconstructed data samples instead of user data for content popularity prediction in the BS.
Assuming the content library contains $F$ items, the dimension of $\tilde{g}_{u}$ is $F$ and can be expressed as $\tilde{g}_{u}(1,2,...,F)$. All reconstructed fake samples can be added by dimension to obtain the score $\tilde{g}(1,2,...,F)$ of all contents,
\begin{equation}
	\tilde{g}(1,2,...,F)=\frac{1}{U}\sum_{u=1}^{U}\tilde{g}_{u}(1,2,...,F).
\end{equation}
The score $\tilde{g}(1,2,...,F)$ reflects the overall preferences of users within the BS coverage area, which does not expose the privacy of individual users. The higher the score, the more popular the content is. Then, considering the cache capacity of the BS, cache the $N$ most popular contents. The above process corresponds to step 5 in Fig. \ref{system model}.

\section{Simulation}
In this section, we conducted experiments on a CPU with a maximum speed of 5.3GHz while using the widely recognized MovieLens 1M dataset. The MovieLens 1M dataset includes 1,002,099 ratings from 6,040 users on 3,952 movies, with each rating ranging from 0 to 5. The values of the parameters in the experiment are shown in Table~\ref{tab1}. Unless otherwise specified, the number of users participating in the training is 20, the time steps $T$ is 50. and the BS cache capacity is 100 contents. 

Our overall U-Net architecture is similar to \cite{UNet}. The main differences are the replacement of 2D convolutions with 1D convolutions to adapt to the user’s interaction data structure, and the use of one-fourth of the number of channels and three feature map resolutions to reduce the model size \cite{grad, diffwave}. Therefore, our LDPM has only 770K parameters, making it suitable for resource-constrained edge devices. 
To evaluate the scheme, we adopt the cache hit percentage and request content delay as evaluation metrics \cite{cache_2017}. The cache hit percentage represents the success rate of directly requesting content from the BS. The more accurate the predicted popular content, the higher the cache hit percentage. When the requested content is stored in the BS, it is regarded as a successful cache; conversely, if the content is not cached in the BS, it is termed a failed cache. The request content delay represents the average delay of all users getting content.
\begin{figure}[htbp]
	\centering
	\includegraphics[scale=0.45]{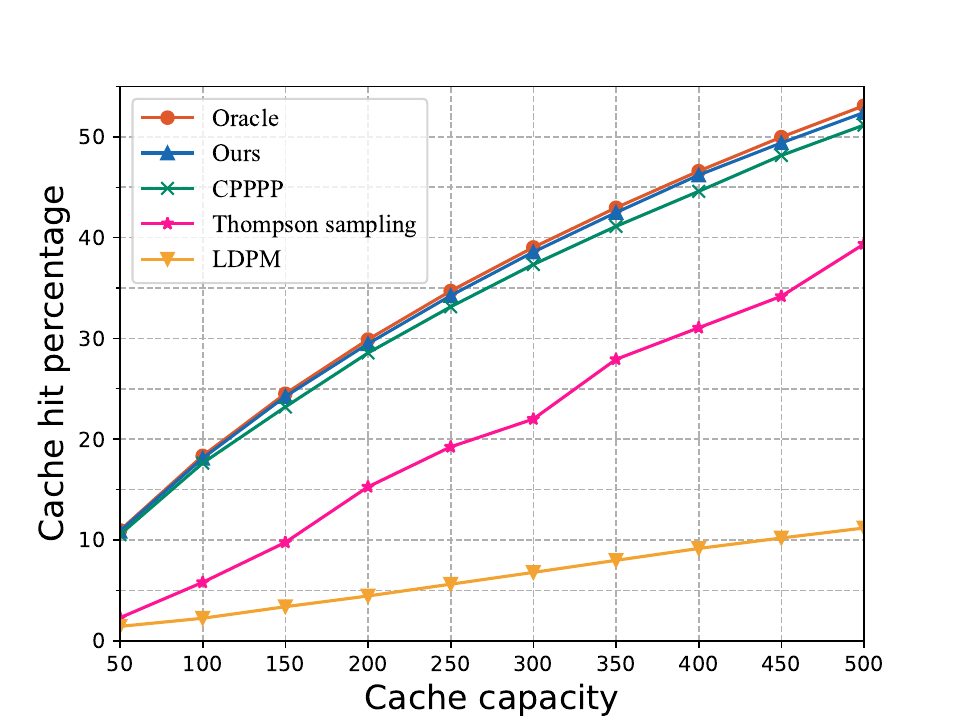}
	\caption{Cache hit percentage versus cache capacity.}
	\label{fig2}
\end{figure}

We compare our proposed edge caching scheme with other schemes, such as:
\begin{itemize}
	\item Oracle \cite{cache_2017}: The Oracle algorithm possesses full prior knowledge of users future requests, defining the theoretical maximum achievable cache hit percentage.
	\item CPPPP \cite{WGAN}: Claims to be the first to use FL with GAN to predict popular content.
	\item Thompson Sampling: In each iteration, the BS dynamically updates its cached contents by evaluating historical cache success/failure statistics and retains the top $N$ highest-value items through Bayesian posterior probability updates.
	\item LDPM: Directly using raw user data for LDPM training without using the pre-trained encoder for data processing.
\end{itemize}

\begin{figure}[htbp]
	\centering
	\includegraphics[scale=0.45]{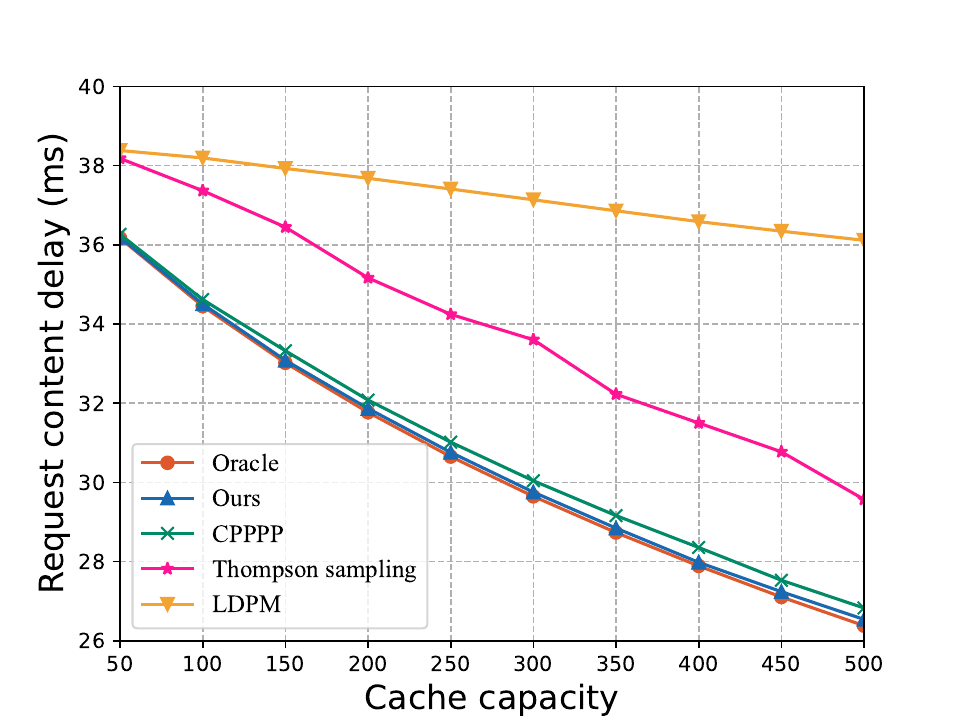}
	\caption{Request content delay versus cache capacity.}
	\label{fig3}
\end{figure}

Fig. \ref{fig2} illustrates the cache hit percentage of BS under different caching capacity across various schemes. It can be observed that as the caching capacity increases, the cache hit percentage improves for all schemes. This is because larger caching capacity enable the BS to store more content, making it more likely for users to retrieve the requested content from the BS. Oracle has the highest cache hit percentage because it knows the content of user requests in the future. Our proposed scheme outperform CPPPP because LDPM leverage a step-by-step denoising generation process, a stable training objective, and a more comprehensive ability to approximate data distributions, effectively overcoming the limitations of GAN in terms of training instability and mode collapse. The performance of CPPPP is superior to Thompson Sampling because Thompson Sampling does not rely on learning-based content prediction. LDPM has the worst performance because it is difficult for LDPM to learn an effective data distribution directly on the original high-dimensional sparse user data.

Fig. \ref{fig3} shows the request content delay of BS under different caching capacity for various schemes. It can be observed that as the caching capacity increases, the request content delay decreases across all schemes. This is because a larger caching capacity allows the BS to store more content, increasing the likelihood that each user can obtain the desired content directly from the BS, thereby reducing the request delay. Furthermore, the request delay of our proposed scheme is lower than that of other schemes except for Oracle. This is attributed to the higher cache hit percentage of our proposed scheme, which enables more users to retrieve content from the BS, further minimizing the request delay.

\begin{table}
	\caption{Values of the parameters in the experiments.}
	\label{tab1}
	\footnotesize
	\centering
	\begin{tabular}{|c|c|c|c|}
		\hline
		\textbf{Parameter} &\textbf{Value} \\
		\hline
		$\eta_d$ & 0.0006  \\
		\hline
		$e$ & 30 \\
		\hline
		$U$ & 1000 \\
		\hline
		$F$ & 3952 \\
		\hline
		Structure of pre-trained encoder & 3952-100-16 \\
		\hline
		Structure of pre-trained decoder & 16-100-3952 \\
		\hline
	\end{tabular}
\end{table}

\begin{table*}[ht]
	\caption{Cache efficiency and total training time under different time steps $T$.}
	\label{tab2}
	\footnotesize
	\centering
	\begin{tabular}{|c|cccccccccc|c|}
		\hline
		\multirow{2}{*}{T}  &  \multicolumn{10}{c|}{Cache capacity}&  \multirow{2}{*}{Total training time (s)} \\ 
		\cline{2-11}
		& 50      & 100     & 150     & 200     & 250     & 300     & 350     & 400     & 450     & 500 & \\  
		\hline
		\textbf{10}                    & 10.18\% & 17.38\% & 22.85\% & 28.22\% & 32.94\% & 37.08\% & 40.81\% & 44.55\% & 47.80\% & 51.01\% & 18.46\\
		\textbf{50}                     & 10.79\% & 17.67\% & 23.67\% & 29.06\% & 33.83\% & 38.23\% & 42.30\% & 45.75\% & 48.98\% & 52.20\% & 21.56\\
		\textbf{100}                  & 10.83\% & 17.95\% & 24.02\% & 29.21\% & 34.09\% & 38.20\% & 42.13\% & 45.68\% & 49.20\% & 52.25\% & 33.88\\
		\textbf{200}                  & 10.93\%  & 18.00\%  & 24.09\%  & 29.37\%  & 34.09\%  & 38.34\%  & 42.31\%  & 45.92\%  & 49.17\% & 52.18\% & 40.68\\
		\textbf{500}      & 10.87\%  & 18.09\%  & 24.20\% & 29.62\% & 34.31\% & 38.53\% & 42.44\% & 46.18\% & 49.38\% & 52.39\% & 92.10\\\hline
	\end{tabular}
\end{table*}

From Table~\ref{tab2}, it can be observed that as the time steps $T$ increases from 10 to 50, the cache hit percentage for different cache capacities show a significant improvement, while the total training time of the model increases slightly. This is because the increase in the number of time steps $T$ allows the model to learn more refined noise, thus producing better performance. Further increasing time steps $T$ results in almost no change in the cache hit percentage for different cache capacities, but significantly increases the total training time of the model. This is because the model has already achieved near-optimal performance, and further increasing the number of time steps $T$ cannot significantly improve performance but will significantly increase the model training time. Therefore, we choose $T = 50$.
\vspace{-0.2cm}
\begin{figure}[htbp]
	\centering
	\includegraphics[scale=0.45]{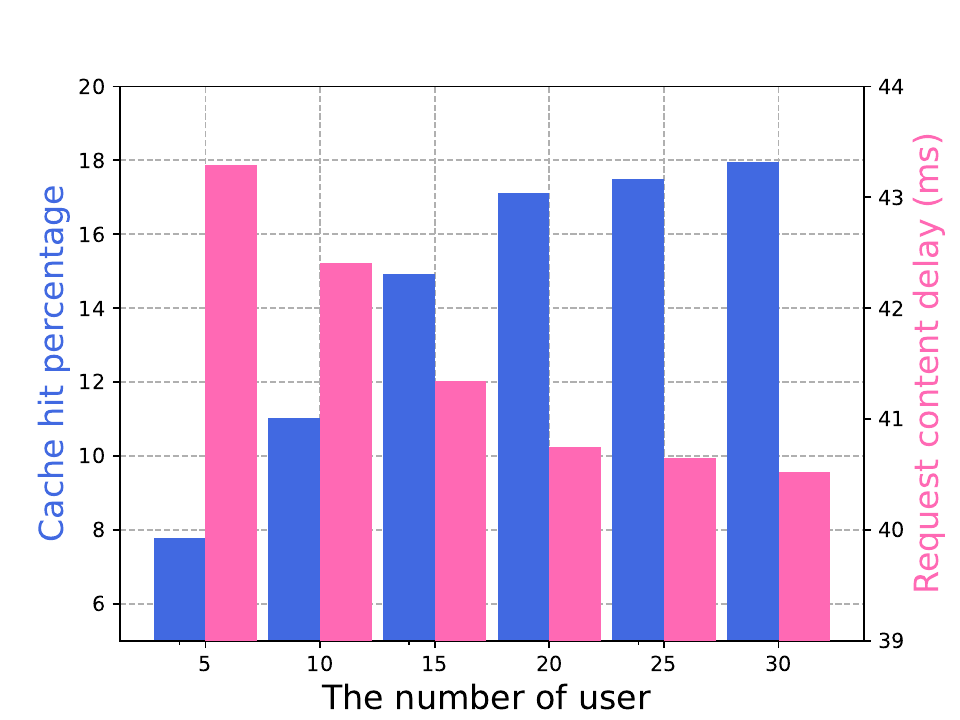}
	\caption{Cache hit percentage and request content delay versus the number of user.}
	\label{fig4}
\end{figure}

Fig. \ref{fig4} shows how cache hit percentage and request content delay vary with the number of user participating in training. It can be seen that as the number of user participating in training increases, the cache hit percentage gradually increases and the request content delay gradually decreases. This is because more users provide more data and computational power, which allows for more accurate prediction of popular content.

\section{Conclusion}
In this letter, we propose a FL assisted edge caching scheme based on LDPM, achieving a higher cache hit percentage compared to existing FL-based methods and baseline methods. To protect user privacy, we first propose a federated-based LDPM training algorithm. Afterwards, we propose an algorithm for predicting popular content on edge nodes. Finally, experiments are conducted to verify the scheme we proposed.

\vfill
	
\end{document}